\documentstyle[12pt]{article} 
\begin{document} 
\topmargin -1.0truecm
\title{ \hfill
hep-ph/9808261\\ \vskip 1.5truecm
{\large{\bf Gravitationally Violated $U(1)$ Symmetry
and}}\\
{\large{\bf Neutrino
Anomalies}}} \vskip 2.0truecm

\author{ Anjan S. Joshipura \\
{\ns\it Theoretical Physics Group, Physical Research Laboratory,}\\
{\ns\it Navarangpura, Ahmedabad, 380 009, India.}}

\date{}
\def\ap#1#2#3{           {\it Ann. Phys. (NY) }{\bf #1} (19#2) #3}
\def\arnps#1#2#3{        {\it Ann. Rev. Nucl. Part. Sci. }{\bf #1} (19#2) #3}
\def\cnpp#1#2#3{        {\it Comm. Nucl. Part. Phys. }{\bf #1} (19#2) #3}
\def\apj#1#2#3{          {\it Astrophys. J. }{\bf #1} (19#2) #3}
\def\asr#1#2#3{          {\it Astrophys. Space Rev. }{\bf #1} (19#2) #3}
\def\ass#1#2#3{          {\it Astrophys. Space Sci. }{\bf #1} (19#2) #3}

\def\apjl#1#2#3{         {\it Astrophys. J. Lett. }{\bf #1} (19#2) #3}
\def\ass#1#2#3{          {\it Astrophys. Space Sci. }{\bf #1} (19#2) #3}
\def\jel#1#2#3{         {\it Journal Europhys. Lett. }{\bf #1} (19#2) #3}

\def\ib#1#2#3{           {\it ibid. }{\bf #1} (19#2) #3}
\def\nat#1#2#3{          {\it Nature }{\bf #1} (19#2) #3}
\def\nps#1#2#3{          {\it Nucl. Phys. B (Proc. Suppl.) }
                         {\bf #1} (19#2) #3} 
\def\np#1#2#3{           {\it Nucl. Phys. }{\bf #1} (19#2) #3}
\def\pl#1#2#3{           {\it Phys. Lett. }{\bf #1} (19#2) #3}
\def\pr#1#2#3{           {\it Phys. Rev. }{\bf #1} (19#2) #3}
\def\prep#1#2#3{         {\it Phys. Rep. }{\bf #1} (19#2) #3}
\def\prl#1#2#3{          {\it Phys. Rev. Lett. }{\bf #1} (19#2) #3}
\def\pw#1#2#3{          {\it Particle World }{\bf #1} (19#2) #3}
\def\ptp#1#2#3{          {\it Prog. Theor. Phys. }{\bf #1} (19#2) #3}
\def\jppnp#1#2#3{         {\it J. Prog. Part. Nucl. Phys. }{\bf #1} (19#2) #3}

\def\rpp#1#2#3{         {\it Rep. on Prog. in Phys. }{\bf #1} (19#2) #3}
\def\ptps#1#2#3{         {\it Prog. Theor. Phys. Suppl. }{\bf #1} (19#2) #3}
\def\rmp#1#2#3{          {\it Rev. Mod. Phys. }{\bf #1} (19#2) #3}
\def\zp#1#2#3{           {\it Zeit. fur Physik }{\bf #1} (19#2) #3}
\def\fp#1#2#3{           {\it Fortschr. Phys. }{\bf #1} (19#2) #3}
\def\Zp#1#2#3{           {\it Z. Physik }{\bf #1} (19#2) #3}
\def\Sci#1#2#3{          {\it Science }{\bf #1} (19#2) #3}
\def\n.c.#1#2#3{         {\it Nuovo Cim. }{\bf #1} (19#2) #3}
\def\r.n.c.#1#2#3{       {\it Riv. del Nuovo Cim. }{\bf #1} (19#2) #3}
\def\sjnp#1#2#3{         {\it Sov. J. Nucl. Phys. }{\bf #1} (19#2) #3}
\def\yf#1#2#3{           {\it Yad. Fiz. }{\bf #1} (19#2) #3}
\def\zetf#1#2#3{         {\it Z. Eksp. Teor. Fiz. }{\bf #1} (19#2) #3}
\def\zetfpr#1#2#3{         {\it Z. Eksp. Teor. Fiz. Pisma. Red. }{\bf #1} (19#2) #3}
\def\jetp#1#2#3{         {\it JETP }{\bf #1} (19#2) #3}
\def\mpl#1#2#3{          {\it Mod. Phys. Lett. }{\bf #1} (19#2) #3}
\def\ufn#1#2#3{          {\it Usp. Fiz. Naut. }{\bf #1} (19#2) #3}
\def\sp#1#2#3{           {\it Sov. Phys.-Usp.}{\bf #1} (19#2) #3}
\def\ppnp#1#2#3{           {\it Prog. Part. Nucl. Phys. }{\bf #1} (19#2) #3}
\def\cnpp#1#2#3{           {\it Comm. Nucl. Part. Phys. }{\bf #1} (19#2) #3}
\def\ijmp#1#2#3{           {\it Int. J. Mod. Phys. }{\bf #1} (19#2) #3}
\def\ic#1#2#3{           {\it Investigaci\'on y Ciencia }{\bf #1} (19#2) #3}
\def\tp{these proceedings}
\def\pc{private communication}
\def\ip{in preparation}
\relax

\newcommand{\GeV}{\,{\rm GeV}}
\newcommand{\MeV}{\,{\rm MeV}}
\newcommand{\keV}{\,{\rm keV}}
\newcommand{\eV}{\,{\rm eV}}
\newcommand{\Tr}{{\rm Tr}\!}
\renewcommand{\arraystretch}{1.2}
\newcommand{\beq}{\begin{equation}}
\newcommand{\eeq}{\end{equation}}
\newcommand{\beqa}{\begin{eqnarray}}
\newcommand{\eeqa}{\end{eqnarray}}
\newcommand{\ba}{\begin{array}}
\newcommand{\ea}{\end{array}}
\newcommand{\bmat}{\left(\ba}
\newcommand{\emat}{\ea\right)}
\newcommand{\refs}[1]{(\ref{#1})}
\newcommand{\ler}{\stackrel{\scriptstyle <}{\scriptstyle\sim}}
\newcommand{\ger}{\stackrel{\scriptstyle >}{\scriptstyle\sim}}
\newcommand{\lag}{\langle}
\newcommand{\rag}{\rangle}
\newcommand{\ns}{\normalsize}
\newcommand{\cm}{{\cal M}}
\newcommand{\gr}{m_{3/2}}
\newcommand{\p}{\partial}

\def\rp{ $R_P$} 
\def\321{$SU(3)\times SU(2)\times U(1)$}
\def\tl{{\tilde{l}}}
\def\tL{{\tilde{L}}}
\def\bd{{\overline{d}}}
\def\tL{{\tilde{L}}}
\def\a{\alpha}
\def\b{\beta}
\def\g{\gamma}
\def\c{\chi}
\def\d{\delta}
\def\D{\Delta}
\def\db{{\overline{\delta}}}
\def\Db{{\overline{\Delta}}}
\def\e{\epsilon}
\def\l{\lambda}
\def\n{\nu}
\def\m{\mu}
\def\nt{{\tilde{\nu}}}
\def\p{\phi}
\def\P{\Phi}
\def\x{\xi}
\def\r{\rho}
\def\s{\sigma}
\def\t{\tau}
\def\th{\theta}
\def\ne{\nu_e}
\def\nm{\nu_{\mu}}
\def\rp{$R_P$}
\def\mp{$M_P$}     
\renewcommand{\Huge}{\Large}
\renewcommand{\LARGE}{\Large}
\renewcommand{\Large}{\large}
\maketitle
\vskip 2.0truecm
\begin{abstract}
The current searches for neutrino oscillations seem to suggest an
approximate $L_e-L_\m-L_{\tau}$ flavor symmetry. This symmetry
implies a pair of degenerate
neutrinos with mass $m_0$  and large  leptonic mixing.
We explore the possibility that gravitational
interactions break this global symmetry. The Planck scale suppressed
breaking of the $L_e-L_\m-L_{\tau}$ symmetry is shown to lead to
the right amount of splitting
among the degenerate neutrinos needed in order to solve  the solar
neutrino problem. The common mass $m_0$ of the pair can be identified with
the 
atmospheric neutrino scale. A concrete model is proposed in which
smallness of $m_0$ and hierarchy in  the solar and atmospheric
neutrino scales get linked to hierarchies in the weak, grand unification 
and the Planck scales.
\end{abstract}
Pattern of neutrino masses and mixing as suggested by the present
experimental evidences and hints seem quite different from the one
in the quark sector. The oscillations of $\nu_\m$ of the atmospheric
origin require large $\nu_\m-\nu_{\tau}$ mixing and very small difference
$\Delta_A\sim 10^{-3} \eV^2$ in their squared masses \cite{superk}. The
solar neutrino
anomalies require \cite{bks} much smaller mass scale $\Delta_S\sim 10^{-6}
\eV^2$
(MSW \cite{msw} conversion) or $\Delta_S\sim 10^{-10}\eV^2$ (vacuum
oscillations \cite{vac}).
The latter alternative can reconcile the solar anomaly only if mixing
involving $\nu_e$ is large. 

The conventional seesaw models based on  grand unified theories 
link the masses and mixing of leptons to that in the quark sector
\cite{pm}.
This link does not seem to be fully supported by the experiments and one
must 
either admit variety of textures \cite{bh,text} in right handed
neutrino
masses or look
for some alternative \cite{alt} schemes. 

The presently available information on the solar neutrinos do not
seem to choose unambiguously \cite{bks} between the MSW or the vacuum
oscillation
solutions although the MSW conversion with large angle seems to be highly
disfavored \cite{bks} by the recent \cite{dayn} day night asymmetry
measurements at
SuperKamioka. Thus the MSW mechanism requires small mixing of $\nu_e$.
It is then intriguing why large mixing is preferred in the
$\nu_\m-\nu_{\tau}$ system with less
hierarchical $\Delta_A$. The vacuum solution seems more natural from the
point of view of this theoretical prejudice but in this case one has 
a problem of accounting  for very large hierarchy ${\Delta_S\over
\Delta_A}\sim
10^{-7}$. This note is devoted to discussion of these issues.

Let us suppose that both the solar and the atmospheric neutrino
oscillations
are described by maximal ($\sim \pi/4$) mixing among relevant states.
This hypothesis is shown \cite{bim} to lead through unitarity to a unique
structure for the mixing matrix U given by:
\beq\label{u}
U=\left( \ba{ccc}
1/\sqrt{2}&-1/\sqrt{2}&0\\
c/\sqrt{2}&c/\sqrt{2}&-s\\
s/\sqrt{2}&s/\sqrt{2}&c\\
\end{array}\right)
\eeq
where, $c=\cos\theta, s=\sin\theta\sim {1\over \sqrt{2}}.$
This structure can describe the solar and atmospheric neutrino
observations successfully if $\Delta_{23}\equiv
m_{\nu_3}^2-m_{\nu_2}^2\sim 10^{-3}\eV^2$ and
$\Delta_{12}\equiv m_{\nu_2}^2-m_{\nu_1}^2\sim 10^{-10} \eV^2$.
It also implies that $\nu_e$ does not oscillate at the atmospheric scale
in
accordance with the findings at SuperKamioka. This $U$ together with
neutrino masses can be used to determine the structure of the 
light neutrino mass matrix in basis with diagonal charged lepton masses.
This was done  \cite{bim} in case of the hierarchical masses. Since 
large mixing may be intimately related to pseudo-Dirac structure, let us
suppose that a pair of neutrinos are (almost) degenerate with masses $ m_0$
and
-$m_0$. This common mass may be identified with the atmospheric neutrino
scale. For a fixed $U$ as given in eq.(\ref{u}), one has three physically
distinct possibilities corresponding to $m_i$ ($i$=1,2,3) values 
$$(a)\;
(m_0,-m_0,0)\;\;\;\; (b)\; (m_0,0,-m_0)
\;\;\;\;(c)\;\;\;\;(0,m_0,-m_0)\;.$$
This implies the following neutrino mass matrices $M_\n$ for the three
light
states:
\beq \label{textures} 
(a)\;\; m_0 \left( \ba{ccc}
0&c&s\\
c&0&0\\
s&0&0\\
\end{array}\right)\;\;\;\;\;\;\;\;\;
(b)\;\;\;{m_0\over 2}
\left( \ba{ccc}
1&c&s\\
c&1-3 s^2&3cs\\
s&3cs&1-3 c^2\\
\end{array}\right) \nonumber \\
\eeq
\beq
(c)\;\;\;{m_0\over 2}
\left( \ba{ccc}
1&-c&-s\\
-c&1-3 s^2&3cs\\
-s&3cs&1-3 c^2\\
\end{array}\right)
\eeq
Of these, the texture in $(a)$ seems more interesting as it does not
presuppose any relations among matrix elements of $M_\n$. Moreover,
this texture follows from a simple $L_e-L_\m-L_{\tau}$ symmetry. 
Conversely, bimaximal mixing may be regarded \cite{bh} as a consequence of 
the $L_e-L_\m-L_{\tau}$ symmetry imposed in the  leptonic
sector \cite{note}.
One still
needs to 
understand the origin of $m_0$ and of much smaller splitting  
$\Delta_S$ between (almost) degenerate pair. The splitting
may arise due to small breaking of the $L_e-L_\m-L_{\tau}$ symmetry.
This can be parameterized \cite{bh} in terms of a small parameter $\e$
leading
to
\beq \label{eps}
M_\n=m_0\left( \ba{ccc}
\e&c&s\\
c&\e&\e\\
s&\e&\e\\
\end{array}\right)
\eeq
where different entries are meant to denote the order of magnitudes of 
the breaking term. This leads to
\beq
\Delta_S\sim 4 m_0 \e \eeq
When $m_0$ is identified with the atmospheric scale ($\sim .03 \eV$), the
above equation
implies
\beqa \label{evalue}
 \e&\sim  (10^{-4}-10^{-5})&\;\;\;\;
{\mbox for\;\;} \Delta_S\sim (10^{-5}-10^{-6})\eV^2 \;({\mbox MSW})
\nonumber
\\
\e&\sim (10^{-9}-10^{-10})&\;\;\;\;{\mbox for}\;\;
\Delta_S\sim (10^{-10}-10^{-11})\eV^2 \;\; ({\mbox Vacuum}) 
\eeqa
What could be the origin of such small values for $m_0$ and $\e$ ?
It is indeed possible to link these scales to the hierarchies among the
known scales namely $M_{weak},M_H\sim M_{GUT}$ and $M_{Planck}$.
 
Let us consider the standard $SU(2)\times U(1) $ model without addition of
any right handed neutrinos. Neutrino masses are generated through 
the following Yukawa couplings when an
$SU(2)$-triplet Higgs field is introduced:
\beq \label{numass}
-{\cal L}_\n={1\over 2} f_{ij}\;L^T_i \Delta L_j +c.c.
\eeq
Here $\Delta$ refers to the $2\times 2$ matrix for the triplet Higgs
field. We have suppressed the Lorentz indices in the above equation.
$i,j$ refer to the generation indices.
We also impose the $L_e-L_\m-L_{\tau}$
symmetry.
Non-zero vacuum expectation value (vev) for the neutral component of
$\Delta$ then leads to
structure as in eq.(\ref{textures}a) with,
$$m_0=(f_{12}^2+f_{13}^2)^{1/2}<\Delta^0>\;\;\; ;\tan\theta={f_{13}\over
f_{12}}.$$ 
$\theta$ could be naturally large for $f_{12}\sim f_{13}$.
The smallness of $m_0$ may appear unnatural in $SU(2)\times U(1)$ theory.
But small triplet vev and hence $m_0$ may result  if theory contains heavy
scales such as $M_{GUT}$. This follows from the induced vev mechanism
which implies \cite{pm,ms}:
$$m_0\sim\;\; <\Delta^0>\;\;\sim {M_W^2\over M_H}.$$    
$M_H\sim 10^{15}\GeV$ then leads  to the required
value $m_0\sim 10^{-2} \eV$. 

The symmetry $L_e-L_\m-L_{\tau}$ must be regarded as a global symmetry
in the present context since it is not possible to gauge this
symmetry
in  standard model (SM) without introducing right handed neutrinos. Such
global
symmetries are known to be unstable against gravitational effects
~\cite{gre1,gre2}
and would be broken.  We
assume that this breaking is manifested in the low
energy theory through higher dimensional  operators suppressed by the 
Planck mass $M_P$. One could write the following
symmetry breaking non-renormalizable terms:
\beqa \label{hdo}
O_1={1\over 2}\beta_{1ij}\nu_i^T \nu_j  ({\phi^{0*2} \over M_P})\nonumber
\\
O_2={1\over 2}\beta_{2ij} \nu_i^T \nu_j\Delta^0 \left({\eta\over M_P}\right) 
\nonumber \\
O_3={1\over 2}\beta_{3ij} \nu_i^T \nu_j \Delta^0 \left({\eta^{\dagger}\eta
\over
M_P^2}\right)
\eeqa     
Here we have introduced $SU(2)\times U(1)$ singlet field $\eta$ which is
assumed to obtain large $\sim M_H$  vev. $\phi^0$ ($\Delta^0$) corresponds
to the neutral component of the 
$SU(2)\times U(1)$ doublet ( triplet ) Higgs field. The dimensionless
couplings $\b_{mij}\;\;(m=1,2,3)$ break the $L_e-L_\m-L_{\tau}$ symmetry.

The operator $O_1$ is the familiar one introduced for example in
\cite{azs}. The above operators lead to the $\e$
parameters
in eq.(\ref{eps}) and
hence to the splitting among the degenerate pair. One respectively gets
for the operators $O_1-O_3$ 
\beqa \label {spliting}
(\Delta m^2)_1\!\!\!&\sim&\!\!\!4 \beta_1m_0 {M_{weak}^2\over M_P}\sim
\beta_1( 10^{-7}\eV^2) \left(
{m_0 \over 10^{-2} \eV}\right) \nonumber \\
(\Delta m^2)_2\!\!\!&\sim&\!\!\!4 \beta_2 m_0^2 \left({<\eta>\over
M_P}\right)\sim\beta_2(4.0\cdot  10^{-6}\eV^2)
\left(
{m_0^2 \over 10^{-3} \eV^2}\right)\left({<\eta>\over 10^{16}\GeV}\right)
\nonumber \\
(\Delta m^2)_3\!\!\!&\sim&\!\!\!4 \beta_3 m_0^2 \left({<\eta>\over
M_P}\right)^2\!
\!\!\sim \!\!
\beta_3(4.0\cdot 10^{-9}\eV^2)\left(
{m_0^2 \over 10^{-3} \eV^2}\right)\left({<\eta>\over
10^{16}\GeV}\right)^2
\eeqa
where $(\Delta m^2)_i$ denote splitting of the degenerate states induced
by $O_i$.
Here we have assumed that parameters $f$ in eq.(\ref{numass}) are  of
O(1) and identified $f <\Delta^0>\sim <\Delta^0>=m_0$. 

The operator $O_1$ gives a splitting which is somewhat larger (smaller)
than the scale needed for the vacuum (MSW) solution to the solar neutrino
problem. The second and the third operators can generate scales relevant
for the MSW and the vacuum solutions respectively if the vev for $\eta$
is at or near the grand unification scale. While MSW is a natural
and appealing   solution to the solar neutrino problem, it cannot be
implemented in the present context for two reasons. Firstly, the
large angle solution obtained here from the $L_e-L_\m-L_{\tau}$
symmetry seems to be disfavored experimentally as already mentioned.
More importantly, the said symmetry implies a mixing angle of $\pi/4$
degree for which the matter effects do not occur. The corrections
to this mixing angle induced due to  $\e$ are too small to change it
appreciably \cite{bh}. Thus in spite of the possibility of generating the
MSW scale
naturally, vacuum solution is to be preferred in the present context.
This solution can be realized easily in a simple model to which we now
turn. 

We extend the SM by adding two additional Higgs fields namely,
an $SU(2)$-triplet $\Delta$ and a singlet $\eta$. In addition, we impose 
$L_e-L_\m-L_{\tau}$ and a
$Z_3$ symmetry with the charge assignment $(1,-1,1,1)$ for the fields
$(u^c,d^c,\phi,\eta)$. Rest of the fields are assumed to carry zero charge
under $Z_3$. All the scalar fields are also assumed to be neutral under 
$L_e-L_\m-L_{\tau}$. The Yukawa couplings in eq.(\ref{numass}) generate
the required $L_e-L_\m-L_{\tau}$ symmetric mass matrix. The smallness
of $\Delta_0$ arises as follows. Consider the following scalar
potential containing a heavy $\sim M_{GUT}$ and the electroweak scale:
\beqa \label{pot}
V&=&\m^2 \phi^{\dagger} \phi+M_{\Delta}^2 Tr.
\Delta^{\dagger}\Delta+M_{\eta}^2 \eta^{\dagger}\eta \nonumber \\
&+&\lambda (\phi^{\dagger} \phi)^2+\lambda_{\Delta} Tr.
(\Delta^{\dagger}\Delta)^2+\lambda_{\eta}(\eta^{\dagger}\eta)^2+....\nonumber
\\
&-&\left[\b \phi^T\Delta\phi\eta+c.c.\right] 
\eeqa
The terms not explicitly written in the above equations correspond to
some of the quartic terms involving $\Delta$ and crossed quartic terms
\cite{f1} for the doublet
field. 
We assume that all the mass scales except the one (namely
$\m^2$) associated with
the $SU(2)$ doublet field in the above equations are large, i.e. $\sim
M_{GUT}$. For $M_{\eta}^2$ negative, the vev for $\eta$ is driven to a
large scale while $\Delta^0$ vev can be small for $M_{\Delta^2}>0$. 
Minimization of eq.(\ref{pot}) gives,
\beqa \label{delta}
u&\sim& -{M_{\eta}^2\over  \lambda_{\eta}}\;\;, \nonumber \\
\omega&\sim& {\b v^2 u\over2  M_{\Delta}^2}\sim
{\b v^2 \over 2 M_H}\;\;,\nonumber \\
v^2&\sim& -{\m^2\over \lambda+{\b^2\over 2\lambda_{\eta}}\left(
{M_{\eta}^2\over M_{\Delta}^2}\right)}\;\;, \eeqa
where, $<\phi^0>\equiv {v\over \sqrt{2}},
<\Delta^0>\equiv {\omega\over \sqrt{2}}, <\eta^0>\equiv {u\over
\sqrt{2}}$.
The
choice $M_H\sim
M_{\Delta}\sim 10^{15}\GeV$ leads to $<\Delta^0>\sim
10^{-2}
\eV$ very close to the atmospheric mass scale $m_0\sim .03 \eV$.

The above scalar
potential has a global symmetry under which $\Delta$ and $\eta$ carry
opposite charges. This symmetry is spontaneously broken and the potential
leads to a Goldstone boson which can be identified with the majoron. The
above potential in fact coincides formally with the one in the  
triplet plus singlet majoron model
\cite{tm}. The majoron arising in this example remains invisible and 
is phenomenologically consistent.

The structure of the $L_e-L_\m-L_{\tau}$
breaking higher dimensional
operator induced by gravitational effect is governed by the gauge
symmetries of the model. It was realized that this is true even if 
the gauge symmetry of the low energy world is a discrete \cite{wk} one.
The gauged
discrete
symmetries may arise  in the low energy theory as a remnant of some
continuous gauge symmetries if
the Higgs fields responsible for its breaking are invariant under a 
discrete sub group. Such discrete symmetries are then required to 
satisfy the discrete anomaly constraints 
\cite{ir}. These constraints
derived in 
\cite{ir} are given for $Z_N$ group as:
\beqa \label{anoamly}
SU(M)^2XZ_N&:\;\;\;\;\;\;\;&\Sigma T_i q_i={1\over 2} p_M N \nonumber \\
Z_N^3 &:\;\;\;\;\;\;\;&\Sigma q_i^3=m N+\delta n N^3/8
\eeqa
where, $\delta=0\;\;(1) $ for $N$ odd (even). The corresponding 
anomalies involving $U(1)$ factors do not impose any significant
restrictions on the low energy theory \cite{ir}.
It is easily verified that the discrete $Z_3$ imposed here indeed
satisfies these constraints with $p_3=p_2=0,m=-1$ in case 
of the
three
fermionic generations. This symmetry may then be imposed as
an additional constraint in deciding the structure of the allowed higher
dimensional terms. One sees that of the three operators in eq.(\ref{hdo})
only dimension six operator is invariant under the $Z_3$. As mentioned in
eq.(\ref{spliting}), this operator can lead to the right splitting between
the 
degenerate pairs to account for the solar neutrino deficit through vacuum
oscillations. It is indeed remarkable that one could relate  both the
solar and the atmospheric scales to the other known scales this way.

The imposition of a discrete symmetry above is somewhat ad-hoc and may be
dispensed with if the coefficient $\b_{1,2}$ associated with dimension
five terms are small instead of being O(1). Specifically, one requires 
$\b_1\sim 10^{-3}\;,\b_2\sim 10^{-4}$ in eq.({\ref{spliting}) in order to
account for the vacuum value for $\Delta_S$. This suppression need not be
as unnatural as
it may look. A familiar example of such suppression \cite{gre2,sup} is
provided in case of
the breaking of the PQ symmetry \cite{axion} induced by the wormhole
effects \cite{gre1}.
It is found that if the global symmetry in question is spontaneously
broken at a scale $f$ then coefficients characterizing its gravitational
breaking are suppressed by the wormhole action. Such suppression
is typically expected \cite{sup} to be  ${f\over M_P}$. Thus in our case,
spontaneous
breaking of the $L_e-L_{\m}-L_{\tau}$ symmetry around the GUT scale may
account for the required suppression in $\b_{1,2}$. 

We have restricted ourselves so far to the SM. Many of the present
considerations can be generalized to the $SU(5)$ model with some
modifications. The triplet
$\Delta$ may be part of a 15 dimensional representation (denoted by the
same symbol) of $SU(5)$ and the
role
of the singlet field may be played by the adjoint ($A$) representation 
used for
breaking the $SU(5)$ symmetry. A straightforward generalization of the
$L_e-L_\m-L_{\tau}$ symmetry would be to assume a family dependent
$U(1)$ symmetry and
assign charges (1,-1,-1) respectively to three generations of the 
${\bar 5}$-plet  of fermions leaving rest of the fields neutral under
it.  The couplings  ${\bar 5}_i{\bar 5}_j\Delta$ then 
lead to the neutrino masses as in eq.(\ref{numass}) if the triplet
component
of the $15$-plet has a small vev. Such vev could follow \cite{pm} from
a term in the scalar potential coupling the 5-dimensional Higgs field
${\bar H}$ to $\Delta$ 
\beq \label{su5}
 \beta {\bar H}_a{\bar H}_b\Delta^{ac}A_c^b. \eeq
where,  $a,b$ refer to the $SU(5)$
indices.
This term is analogous to the last term in eq.(\ref{pot}).
As in that case, the vevs for the doublet component of ${\bar H}$ and
the adjoint field  induce a vev for the triplet in 15.

The splitting among neutrinos is accounted for by the following
dim 5 operators:
\beq
{\b_{1ij}\over 2 M_P} {\bar 5}_{ia}{\bar 5}_{jb}{ H}^a {
H}^b\;\;\;\;
{\b_{2ij}\over 2 M_P} {\bar 5}_{ia}{\bar 5}_{jb}\Delta^{ac}A_c^b
\eeq

These are analogous to operators $O_{1,2}$ in eq.(\ref{hdo}) and can
account
for the vacuum oscillation scale provided the coefficients
$\beta_{1,2}$ are suppressed.

In the exact $U(1)$ symmetric limit, the down quarks remain massless
while the mass matrix for the up-quark is not restricted by the imposed
$U(1)$ symmetry. The former can
obtain masses from the $U(1)$ breaking terms. These are
characterized by the following  dimension five operators
\beq \label{down}
{\Gamma^d_{ij} \over M_P}{\bar 5}_{ai}10^{ab}_j H_c A^c_b
\eeq
 The adjoint field will acquire  a vev at the grand unification
scale
$$<A^a_a>\sim {M_{X}\over g_{GUT}}$$
where $M_X$ is mass of the $SU(5)$ gauge Boson. For $M_X\sim 10^{16}$
the above operator leads to a contribution of $\leq$ O(0.1 GeV) which is
right
for the description of the strange quark mass but falls short of the value
of the $b$ quark mass. 

Let us consider an alternative possibility in which one assigns
non-trivial $U(1)$-charges also to Higgs fields and the 10-plets of
fermions.
Take as
an example the $U(1)$ assignment $(0,-1/3,-1/3)$ for three 10-plets. The
${\bar 5}$ of Higgs field ${\bar H}$ and the adjoint are assumed
respectively to carry the charges $-2/3$ and $4/3$. The ${\bar H}$ charge 
is
specifically chosen to obtain the right structure for the quark masses.
The  charge for $A$ is fixed by requiring that the term in eq.(\ref{su5})
be allowed by the $U(1)$.
One now obtains
the following mass matrices for the up and down quarks in the absence of
the gravitational breaking of the symmetry:
\beq
M_d=\left( \ba{ccc}
0&m_1'&m_2'\\
0&0&0\\
0&0&0\\ \ea\right)\;\;\;\;\;\;   
M_u=\left( \ba{ccc}
0&0&0\\
0&m_1&m_2\\
0&m_2&m_3\\ \ea \right) \eeq
where $m_{1,2}'$ ($m_{1,2,3}$) are parameters determining down  
(up-quark) masses. It is seen
that the $b,c$ and $t$ quarks acquire masses at this stage. The 
higher dimensional terms displayed in eq.(\ref{down}) can now account for 
the strange and down quark masses. Similarly, one could write dimension 5
operator analogous to eq.(\ref{down}) giving mass to the up-quark.
Thus a large part of quark masses and mixings may actually be due 
to the gravitational breaking of the $U(1)$ symmetry. 
This symmetry  is however not strong enough to
make definitive predictions on theses masses and mixings.

The symmetry $U(1)$ does not remain exact in the last example but is
spontaneously broken around the GUT scale. This may be a welcome feature
as such
breaking can possibly account for  suppression \cite{sup} in 
the magnitudes of the coefficient $\b_{1,2}$ of the higher dimensional
term.

In summary, we have underlined the role that the $L_{e}-L_{\m}-L_{\tau}$
symmetry can play in generating leptonic mixing structure desired on
experimental grounds. The presence of a heavy scale $M_H$ in theory 
then accounts for the atmospheric mass scale. Planck scale suppressed
breaking
of the symmetry seems to be in the correct range to provide a solution to
the solar anomaly as well. The role such breaking can play in generation
of
neutrino masses has been emphasized previously \cite{azs}. Here we have
shown that the Planck scale alongwith $M_{weak}$ and a $M_H\sim
M_{GUT}$ can
account for all the observed features of the solar and atmospheric
anomalies 
provided neutrino mass structure is approximately
$L_{e}-L_{\m}-L_{\tau}$
symmetric.

{\bf Acknowledgments:} I am grateful to Probir Roy, Saurabh Rindani and
Sudhir Vempati
for discussions.


\begin{thebibliography}{99}
\bibitem{note} Note that the $L_e-L_\m-L_{\tau}$ symmetry allows a
non-diagonal charged lepton mass matrix mixing $\mu$ and $\tau$ with some
angle
$\theta_l$. The effect of this is to replace the angle $\theta$ in
eq.(\ref{u}) by $\theta-\theta_l$. This combination needs to be large.
\bibitem{f1} Note that parameters of some of these quartic terms
would need to be fine tunned in order to obtain  the doublet vev at the
electroweak scale. This fine tunning is a reflection of the
familier gauge hierarchy problem and can be avoided by 
considering a supersymmetric version of the model.
\end{thebibliography}
\end{document}